# Fine-Tuning Open-Source Large Language Models to Improve Their Performance on Radiation Oncology Tasks: A Feasibility Study to Investigate Their Potential Clinical Applications in Radiation Oncology

Peilong Wang, PhD[1*], Zhengliang Liu, MS[2*], Yiwei Li, MS[2], Jason Holmes, PhD[1], Peng Shu, MS[2], Lian Zhang, PhD[1], Xiang Li, PhD[3], Quanzheng Li, PhD[3], Brady S. Laughlin, MD[1], Diego Santos Toesca, MD[1], Sujay A. Vora, MD[1], Samir H. Patel, MD[1], Terence T. Sio, MD[1], Tianming Liu, PhD[2], Wei Liu, PhD[2]

[1]Department of Radiation Oncology, Mayo Clinic Arizona, Phoenix, AZ 85054, USA
[2]School of Computing, University of Georgia, Athens, GA 30602, USA
[3]Department of Radiology, Massachusetts General Hospital and Harvard Medical School, Boston, MA 02114, USA

[*]Co-first authors who contribute to this paper equally

Corresponding author: Wei Liu, PhD, Professor of Radiation Oncology, Department of Radiation Oncology, Mayo Clinic Arizona; e-mail: Liu.Wei@mayo.edu.

**Abstract**

**Background**: The radiation oncology clinical practice involves many steps relying on the dynamic interplay of abundant text data. Large language models have displayed remarkable capabilities in processing complex text information. But their direct applications in specific fields like radiation oncology remain underexplored.

**Purpose**: This study aims to investigate whether fine-tuning LLMs with domain knowledge can improve the performance on Task (1) treatment regimen generation, Task (2) treatment modality selection (photon, proton, electron, or brachytherapy), and Task (3) ICD-10 code prediction in radiation oncology.

**Methods**: Data for 15,724 patient cases were extracted. Cases where patients had a single diagnostic record, and a clearly identifiable primary treatment plan were selected for preprocessing and manual annotation to have 7,903 cases of the patient diagnosis, treatment plan, treatment modality, and ICD-10 code. Each case was used to construct a pair consisting of patient diagnostics details and an answer (treatment regimen, treatment modality, or ICD-10 code respectively) for the supervised fine-tuning of these three tasks. Open source LLaMA2-7B and Mistral-7B models were utilized for the fine-tuning with the Low-Rank Approximations method. Accuracy and ROUGE-1 score were reported for the fine-tuned models and original models. Clinical evaluation was performed on Task (1) by radiation oncologists, while precision, recall, and F-1 score were evaluated for Task (2) and (3). One-sided Wilcoxon signed-rank tests were used to statistically analyze the results.

**Results**: Fine-tuned LLMs outperformed original LLMs across all tasks with p-value <= 0.001. Clinical evaluation demonstrated that over 60% of the fine-tuned LLMs-generated treatment regimens were clinically acceptable. Precision, recall, and F1-score showed improved performance of fine-tuned LLMs.

**Conclusions**: Fine-tuned LLMs demonstrated statistically significant improvements over original LLMs upon three clinically important tasks in radiation oncology. This study explored the feasibility of applying fine-tuned LLMs in radiation oncology, inspiring further development of utilizing LLMs to assist with radiation oncology tasks.

## Introduction

The radiation oncology clinical practice presents a high level of complexity and stringent requirements for precision[1, 2]. It involves sequential steps like consultation, simulation, planning, quality assurance, treatment delivery, and patient follow-up[3, 4], relying on the dynamic interplay between abundant text and imaging data. This intricate process is traditionally time-consuming, dependent on manual analysis of vast amounts of unstructured clinical data, and susceptible to variations in human interpretation.

Large Language Models (LLMs) such as ChatGPT[5] have displayed remarkable capabilities in natural language processing (NLP) on many topics[6-8]. However, their direct application in specific domains like healthcare has posed challenges. Specifically, in radiation oncology, a sector demanding a lot of clinical experience and utmost precision, the generic nature of mainstream LLMs like ChatGPT falls short. This is especially true for advanced radiation therapy techniques requiring even more complexity and higher precision, such as Intensity-Modulated Proton Therapy (IMPT)[9-13]. Moreover, medical institutes and healthcare practitioners may need to have their localized LLMs because of the privacy regulations for patient health information (PHI). Additionally, the application of NLP in radiation oncology, although is expanding, remains underexplored, while there are enormous amounts of text and image data available in this field.

As efficient tools for language-involved processing can significantly enhance each phase of radiation therapy and potentially improve treatment outcomes, the necessity arises for a model adapted to clinical domain knowledge with the conciseness and specificity inherent in radiation oncology. Therefore, we finetuned open source LLMs to improve their performance on radiation oncology tasks. Particularly, we focus on the tasks of (1) generating radiotherapy treatment regimens, (2) determining radiation treatment modality (photon, proton, electron, or brachytherapy), and (3) predicting ICD-10 codes[a] based on patient diagnosis details. To our knowledge, this represents the first exploration of fine-tuned LLMs specifically for these radiation oncology tasks.

## Materials and Methods

### Data extraction

Patient data was extracted using an in-house search engine, running locally in the Department of Radiation Oncology at XXXX, which allows for searching and extracting structured data from the Aria database (ver.15.6) (Varian Medical Systems, USA). Two datasets were created from the database: one containing patient diagnosis details, including the patient ID, ICD codes, disease stages, free-form textual diagnosis details, etc., and the other comprising treatment plan delivery details, including the course ID, plan types[b], radiation modality, intent, and status, the number of treatment fields, the number of fractions, prescription dose details, etc. In total, we extracted 15,724 patient cases (Figure 1).

---

[a] ICD stands for "International Classification of Diseases," ICD code is a unique alphanumeric code used to represent specific diseases, medical conditions, and health-related issues.
[b] Treatment plan type includes primary treatment plan, alternative primary treatment plan, replan, boost plan, etc.

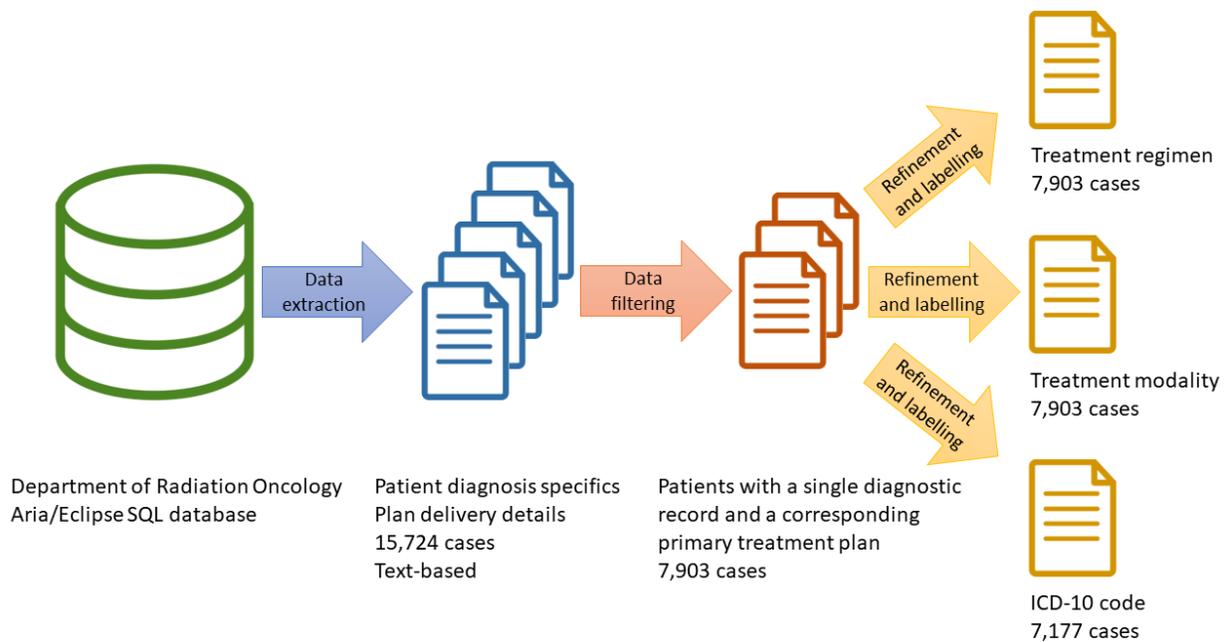

Figure 1. The data extraction and curation process resulted in 7,903 cases for the treatment regime generation and treatment modality selection, and 7,177 cases for ICD-10 code prediction tasks (ICD-9 cases were removed).

Data curation

As high-quality data notably improves LLM fine-tuning results[14], we selected cases where patients only had a single diagnostic record and a clearly identifiable primary treatment plan[c]. For example, data for patients who had a primary treatment plan and an alternative primary treatment plan were removed. This ensured that for each patient diagnosis, there would be only one corresponding primary treatment plan. In total, we selected 7,903 patient cases for the supervised fine-tuning from the 15,724 patients.

---

[c] A primary treatment plan and the subsequent boost treatment plans were included in the dataset.

In some cases, the free-form diagnosis details also contained the treatment plan prescription, for example "IMPT, 35 fractions, 60 Gy", however the patient may have received a slightly different one. Because we wanted to use the treatment planning information corresponding with the treatment plan that was ultimately delivered (based on the ones pulled directly from Aria), we decided to remove any planning information from the diagnosis details (history, pathology, stages). To clean up, we performed a manual process. First, we used a simple keyword filter, primarily searching for the term "Plan" or similar keywords. This successfully identified about 3,000 cases. Subsequently, we manually separated the patient diagnosis and treatment for the remaining 5,000 cases. Lastly, we manually corrected all separate records of diagnosis and treatment. These steps were performed successfully on all 7,903 cases. Additionally, we removed unusual punctuation and initials during this manual correction process to further enhance the quality of the annotated data.

To prepare the annotated treatment modality data, we first merged the patient diagnosis dataset with the plan delivery dataset. Then we selected the treatment modality information from the merged dataset, and standardized the terms as "photon", "proton", "electron" or "brachy". We manually annotated the treatment modality for the cases where this information wasn't identified successfully from the merged dataset, based on the patient diagnosis.

The ICD codes were well-documented within the diagnostic notes, making them directly usable without intricate preprocessing. To ensure data quality, only cases using the ICD-10 standard were kept, and cases using ICD-9 were removed from the selected data. This resulted in 7,177 annotated cases where the patient diagnosis was linked to the correct ICD-10 code.

## Data cohort

The 7,903 selected cases include patients with a mean age of 62 years (range, 1 to 100), with males comprising 56% and females comprising 44%. Regarding the treatment intent, 99% of the patients underwent curative therapy, while the remaining 1% of patients received palliative therapy. The treatment modalities utilized in the dataset are as follows: 64% of patients were treated with photon therapy, 32% received proton therapy, 3% underwent electron therapy, and 1% were treated with brachytherapy.

## Supervised fine-tuning

Fine-tuning is an essential process to tailor pre-trained LLMs for radiation oncology tasks, enabling them to respond more relevantly to these tasks. We utilized LLaMA2-7B[14] (Meta, USA) and Mistral-7B[15] (Mistral AI, France) as the foundational models and used patient diagnosis details as input for the supervised fine-tuning of the three tasks in radiation oncology. We utilized the Low-Rank Approximations (LoRA) method[16] in our supervised fine-tuning, which freezes the pre-trained model weights and injects trainable rank decomposition metrics into each layer of the Transformer architecture.

For the generation of radiotherapy treatment regimens, we constructed 7,903 pairs of input prompts and answers from the 7,903 cases of annotated patient diagnoses and treatment regimens[d]. We engineered the instruction prompt for fine-tuning (Supplementary 1.1) and then fine-tuned LLaMA2 and Mistral models individually on the treatment regimen input prompts and answers (Figure 2). For the selection of radiation treatment modality, we constructed 7,903 pairs of input prompts and treatment modality answers and finetuned the LLaMA2 and Mistral individually on the input prompts and answers in a similar approach. For the prediction of ICD-10 code, 7,177 pairs of input prompts and answers were constructed from the cases of annotated patient diagnosis and ICD-10 code data for fine-tuning. The details of the finetuning parameters are in the Supplementary 1.2. The same configuration parameters were utilized for the fine-tuning of both LLaMA2 and Mistral models.

---
[d] The treatment regimen contains the treatment plan summary given by the radiation oncologist (which may include the drug, dose and fraction, anatomy, etc.) and also the dose prescription summary in the plan delivery notes.

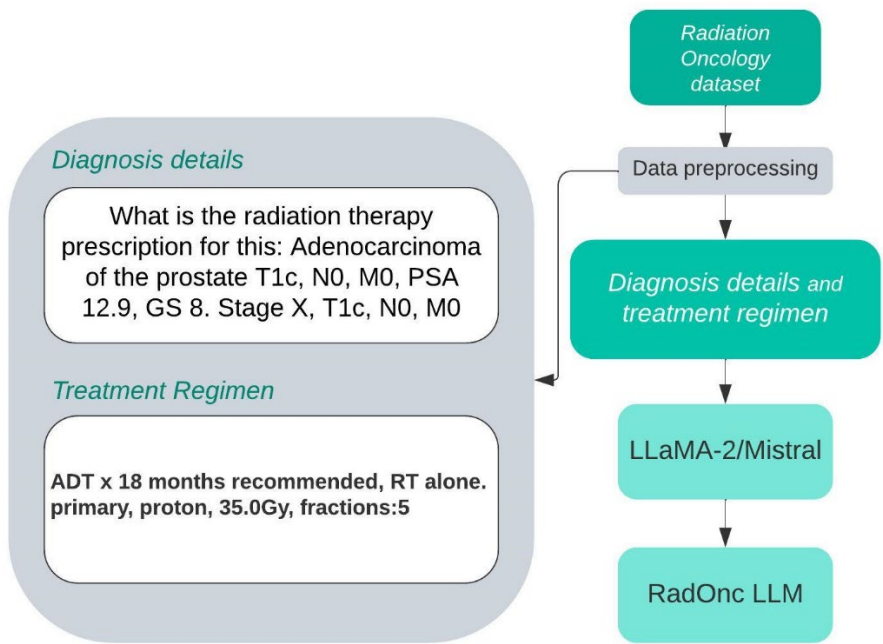

Figure 2. Illustration of the workflow of fine-tuning open-source LLMs for the treatment regimen generation task in radiation oncology. The gray box on the left shows a pair of input prompt and answer used for the open-source LLMs fine-tuning.

For Task (1) and (2), 7,500 cases were used for training and validation, while 403 cases were set aside for independent evaluation for both the vanilla (original) and fine-tuned models. For Task (3), 6,800 cases were used for training and validation, and 377 cases were reserved for independent evaluation (for both vanilla and fine-tuned models). All data were randomized to ensure robust fine-tuning, preventing biases towards specific aspects. The temperature of LLMs was scanned from 0.1 to 1, and statistical analyses were used to check the results. Two NVIDIA A100 80GB GPUs were used for the fine-tuning processes.

Metrics and evaluation

The fine-tuning results are reported in the ROUGE-1 score[e] for Task (1), and accuracy for Task (2) and Task (3). Furthermore, the results of Task (1) were evaluated clinically by our radiation oncologist and medical physicists to check their practical utility in the clinic. Two radiation oncology residents (PGY3 and PGY5) first independently graded the generated treatment regimens with a scale from 1 to 5 based on the following definition:

1. Modality (proton/photon) correct, dose fractionation (SBRT/regular/hypofractionation) correct, clinically completely acceptable and interchangeable.

2. Minor deviation from the physician's plan, by either modality or dose fractionation intent, but still clinically acceptable.

3. More significant deviation from the physician's plan, usually by both modality and dose fractionation intent, potentially unsafe for the patient.

4. Major deviation/errors from the physician's plan, clinically unacceptable/unsafe.

5. Absolutely clinically not acceptable.

Ten grading examples were given by a senior radiation oncologist (15 years of experience) to the residents as references. After grading, cases where the two gradings differed by more than one point were flagged. For these cases, the two residents discussed their gradings with the senior radiation oncologist and an experienced medical physicist to resolve the discrepancies and adjust their scores accordingly.

---

[e] ROUGE-1 = $\frac{\text{Number of overlapping unigrams}}{\text{Total number of unigrams in the reference summary}}$. ROUGE-1 score measures the precision of unigrams (single words) overlap between an automatically generated summary and a reference summary.

For the evaluation of the modality selection and ICD-10 code prediction, precision, recall, and F1 score of the results were used. In addition, confusion matrices were used for the predicted ICD-10 codes. The confusion matrices were made by grouping ICD-10 codes of 377 independent evaluation cases into specific categories based on their descriptions from the official ICD-10 guide. The categories used were "Diseases of the Blood and Blood-Forming Organs," "Congenital Malformations," "Diseases of the Nervous System," "Diseases of the Eye and Adnexa," "Benign Neoplasms," "In Situ Neoplasms," and "Malignant Neoplasms." For the category of "Malignant Neoplasms," further categorization by specific site is conducted. Codes that did not fit into these defined categories were labeled as "Others."

## Statistical analysis

The one-sided Wilcoxon signed-rank test was utilized to compare the performance of the fine-tuned LLMs with those of the vanilla LLMs across three clinical tasks. The null hypothesis suggested no improvement in performance, while the alternative hypothesis anticipated an improvement with the fine-tuned model. Furthermore, we evaluated the performance of LLaMA2 compared with Mistral models using one-sided Wilcoxon signed-rank tests to provide additional insightful findings. All the tests were executed with a 95% confidence interval, based on a significance level of $\alpha = 0.05$. $p$-value smaller than 0.05 is considered to be statistically significant.

**Results:**

Our results showed that the fine-tuned LLMs outperformed the vanilla LLMs across all three tasks (Figure 3). Our clinical evaluation of the treatment regimen generation task demonstrated that over 60% of the treatment regimens generated by the fine-tuned LLMs are clinically acceptable, inspiring future work on further fine-tuning more advanced LLMs for radiation oncology. The details of our findings are reported below.

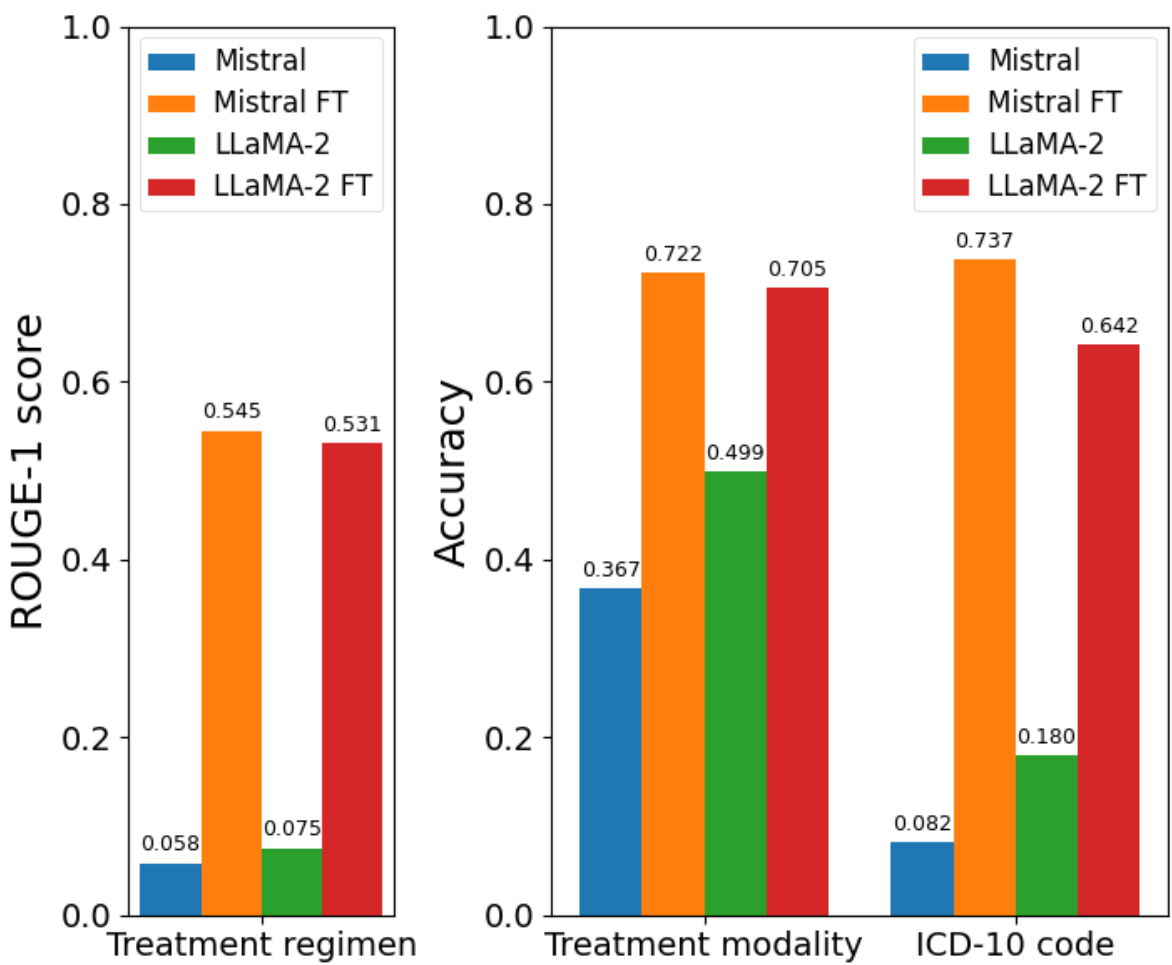

Figure 3. Results of fine-tuned open-source LLMs on treatment regimen generation, treatment modality selection, and ICD-10 code prediction tasks. The performance of fine-tuned LLMs is compared with that of the vanilla ones.

Treatment regimen generation results

In Figure 3, the fine-tuned LLaMA2 model achieved a highest ROUGE-1 score of 0.531 (median: 0.489) after a scan of temperature[f] values ranging from 0.1 to 1.0 with increments of 0.1 (detailed results in Supplementary Figure 1), compared to the vanilla LLaMA2 model's highest ROUGE-1 score of 0.075 (median: 0.072). In Table 1, the one-sided Wilcoxon signed-rank test results indicated a statistically significant increase in ROUGE-1 score for the treatment regimens generated using the fine-tuned LLaMA2 model, compared to the vanilla LLaMA2 model, with a p-value of 0.001 (n = 10). Figure 4 illustrates an example of the treatment regimen generated by the fine-tuned LLaMA2 model, compared with the ones of vanilla model and our physicians.

Similar results were observed for the fine-tuned Mistral model with a p-value of 0.001 (n = 10). Further results of the comparison of LLaMA2 and Mistral model (Vanilla and fine-tuned) can be found in the supplementary Table 2.

Radiation modality selection results

Fine-tuned LLaMA2 model achieved the highest accuracy score of 0.705 (median: 0.676) after a scan of temperature values, compared to the vanilla model's highest accuracy score of 0.499 (median: 0.469). A statistically significant increase in accuracy score was observed for the fine-tuned LLaMA2 model compared to the vanilla model, with a p-value of 0.001 (n = 10). Similar results were observed for the fine-tuned Mistral model.

---

[f] Temperature (T) is defined as a parameter in the output probability $P_i = \frac{e^{\frac{y_i}{T}}}{\sum_{k=1}^{n} e^{\frac{y_k}{T}}}$ to change the output distribution of the model.

Table 1. Summary of the one-sided Wilcoxon signed-rank test results on the performance of the fine-tuned models vs. the vanilla models.

| Fine-tuned (FT) model | vs. | Vanilla model | Tasks | ROUGE-1 or Accuracy, highest (median) | p-value | n |
|---|---|---|---|---|---|---|
| LLaMA2 FT | > | LLaMA2 | Treatment regime | 0.531 (0.489) vs 0.075 (0.072) | 0.001 | 10 |
| | | | Treatment modality | 0.705 (0.676) vs 0.499 (0.469) | 0.001 | 10 |
| | | | ICD-10 code | 0.642 (0.613) vs 0.180 (0.139) | 0.001 | 10 |
| | | | Overall | - | << 0.001 | 30 |
| Mistral FT | > | Mistral | Treatment regime | 0.545 (0.507) vs 0.058 (0.057) | 0.001 | 10 |
| | | | Treatment modality | 0.722 (0.691) vs 0.367 (0.339) | 0.001 | 10 |
| | | | ICD-10 code | 0.737 (0.721) vs 0.082 (0.053) | 0.001 | 10 |
| | | | Overall | - | << 0.001 | 30 |

**Input question**

What is the radiation therapy prescription for this: AJCC 8th edition prognostic stage I (pT1a Nx M0) LEFT invasive lobular carcinoma, grade 2, estrogen receptor positive, progesterone receptor positive, HER2/Neu negative. Stage X,T1a, pN0, M0, G2, HER2 Neg, ER Pos, PR P

**Output from Mayo Clinic (AZ) physician**

plan 25 Gy/ 5 fx photons. primary, photon, 25.0Gy, fractions:5

**Output from fine-tuned LLaMA 2 7B**

Plan 25 Gy/ 5 fx to the left breast with adjuvant chemo. Primary, photon, 25.0Gy, fractions:5

**Output from vanilla LLaMA 2 7B**

The radiation therapy prescription for a patient with left invasive lobular carcinoma (ILC) in the AJCC 8th edition prognostic stage I (pT1a Nx M0), grade 2, estrogen receptor positive (ER+), progesterone receptor positive (PR+), HER2/Neu negative, with Stage X, T1a, pN0, M0, G2, HER2 Neg, ER Pos, PR Pos, would likely involve the following radiation therapy regimen:
1. Radiation Field: The radiation field will include the primary tumor site (T1a) and any involved lymph nodes (pN0). The field may also include surrounding tissues that are at risk of being involved by cancer cells, such as the breast tissue around the tumor.
2. Radiation Dose: The radiation dose will depend on various factors, including the size and location of the tumor, the patient's overall health status, and the desired outcome (e.g., tumor control, minimization of side effects). In general, the dose for stage I breast cancer is in the range of 40-50 Gy in 10-14 fractions.
3. Radiation Technique: The radiation technique used will depend on ...

Figure 4. An example of the treatment regimen generation result by the fine-tuned LLaMA2 model, compared with the one generated by the vanilla model and the one from the physician (similar results were observed for the Mistral model).

### ICD-10 code prediction results

For simplification purposes, only the main category of the ICD-10 code was included in the accuracy calculation and the subcategory was ignored. Fine-tuned LLaMA2 model achieved the highest accuracy score of 0.642 (median: 0.613) after a scan of temperature values, while the vanilla model achieved the highest accuracy score of 0.180 (median: 0.139). A statistically significant increase was observed for the fine-tuned LLaMA2 model compared to the vanilla model, with a p-value of 0.001 (n = 10). Similar results were observed for the fine-tuned Mistral model.

In combination with all three tasks, the fine-tuned LLaMA2 model was observed to outperform the vanilla model statistically significantly in accuracy and ROUGE-1 score, with p-value << 0.001 (n = 30). Similar results were observed for the fine-tuned Mistral model (p-value << 0.001, n = 30). These results indicate that open-source large language models, fine-tuned with the radiation oncology domain-specific data, will improve their performance on these three radiation oncology tasks.

### Clinical evaluation of the treatment regimen generation results

In the clinical evaluation of Task (1), an average of 187 cases (46%) received Grade 1, indicating clinically complete acceptability and interchangeability, for the fine-tuned LLaMA2 model (Figure 5a). Similarly, an average of 208 cases (52%) scored Grade 1 for the fine-tuned Mistral model (Figure 5b).

When considering both clinically acceptable grades ("Grade 1 + Grade 2"), the fine-tuned LLaMA2 model achieved an average of 257 cases (64%) while the fine-tuned Mistral model achieved an average of 265 cases (66%). Given the high degree of personalization in radiotherapy treatment planning, due to factors such as age, underlying diseases, insurance, and personal choices, generating clinically acceptable treatment regimens that are clinically acceptable for over 60% of patients included in the independent evaluation with real-world scenarios based on only short disease descriptions and stage information is remarkable (notwithstanding that only 7B version of LLaMA2 and Mistral models were used for this feasibility study). We did not evaluate the treatment regimen results from the vanilla LLaMA2 and Mistral models due to their high error rates and messy output (examples in the supplementary 1.3), which made them difficult for our residents to grade them accurately.

In the treatment modality selection evaluation, the comparison of weighted average precision, recall and F1 score values before and after fine-tuning revealed improved results for both LLaMA2 and Mistral models, as illustrated in Figure 5c (detailed results in the supplementary Table 3). Since the treatment modality selection for patients can also be personal, given that our model input only consists of simply diagnosis and staging information, achieving precision and recall scores over 70% is considered very good.

In the ICD-10 code prediction evaluation, the confusion matrices (Figure 6a) demonstrated excellent alignment between the finetuned models' output and actual ICD-10 codes of that category, in comparison with the poor prediction from the vanilla models (Figure 6b). The evaluation of ICD-10 code prediction is also shown in Figure 5a using the macro average precision, recall, and F1 score (details in the supplementary Table 4).

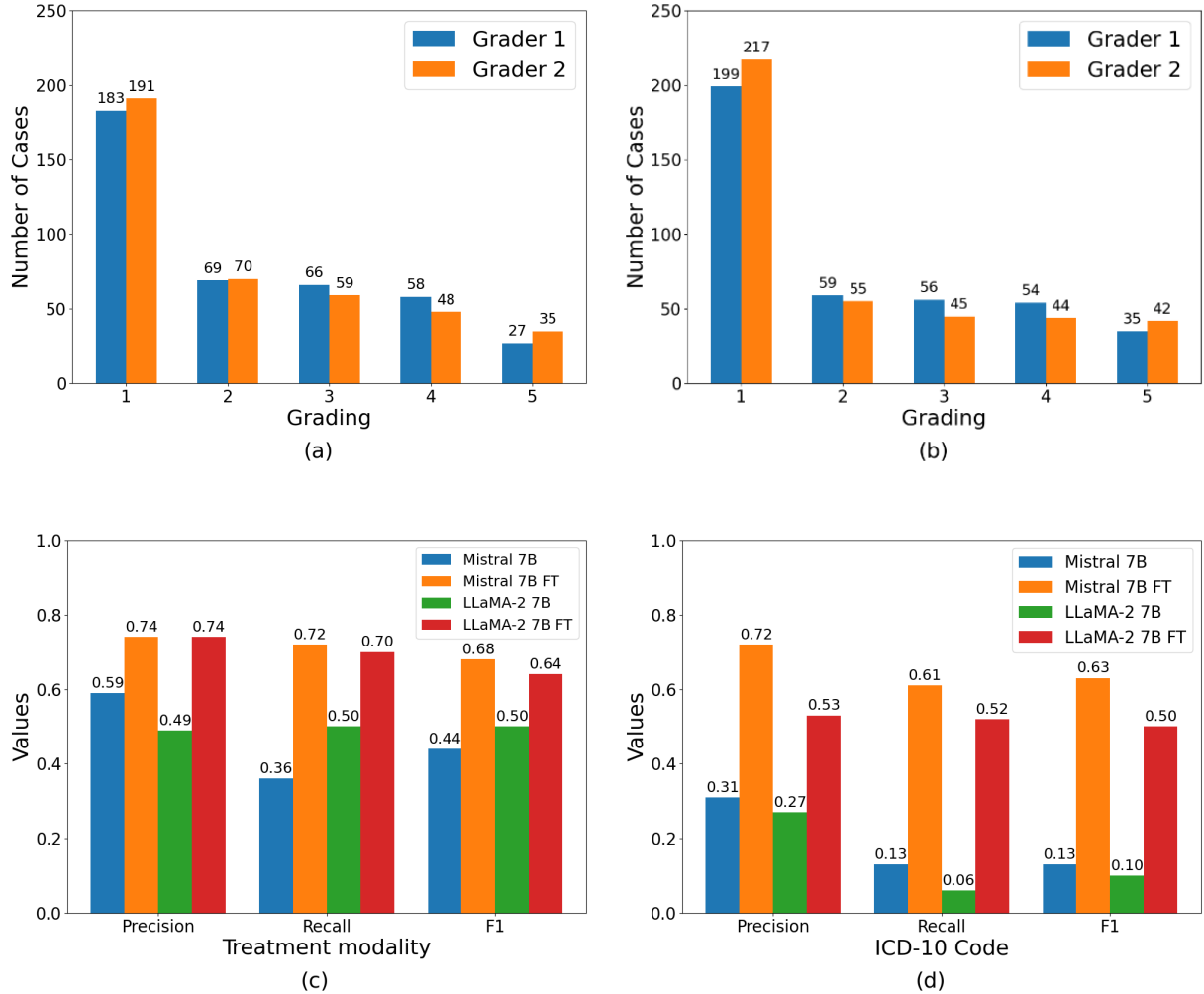

Figure 5. Figure (a) shows the distribution of case gradings for the fine-tuned LLaMA2 model on treatment regimen generation by two graders. Grade 1 is clinically completely acceptable and interchangeable; Grade 5 is absolutely clinically not acceptable. Figure (b) shows the distribution of case gradings for the fine-tuned Mistral model similarly. Figure (c) visualizes the performance of the finetuned and vanilla LLMs on the treatment modality selection using the Precision, Recall, and F1 Score. Figure (d) compares the performance of the finetuned and vanilla models on the ICD-10 code prediction using the Precision, Recall and F1 Score.

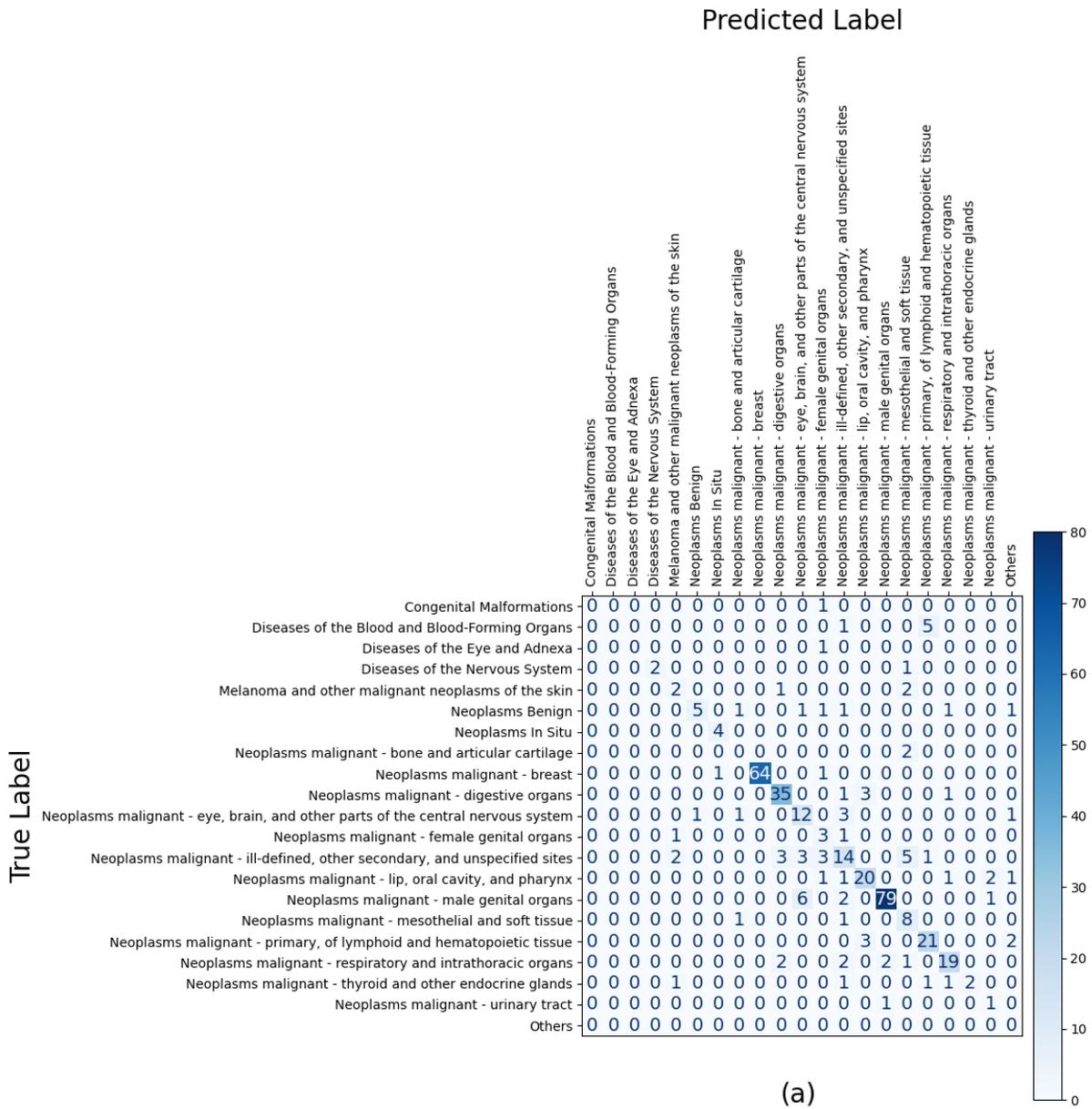

(a)

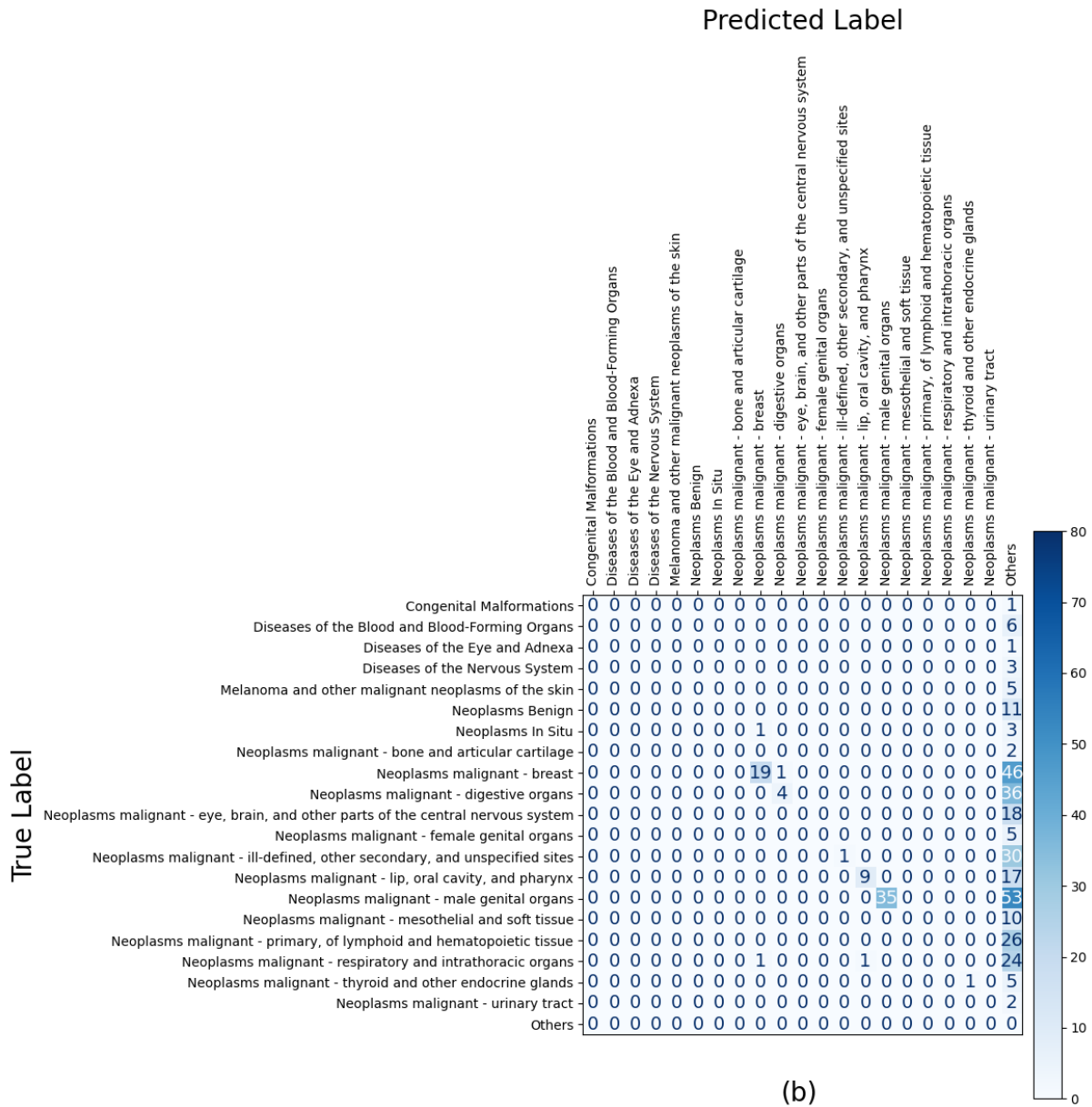

Figure 6. Confusion matrix (a) compares the ICD-10 codes generated by the fine-tuned LLaMA2 with the ICD-10 codes from ground truth. ICD-10 codes are grouped into specific categories for the 377 independent evaluation cases based on the descriptions from the official ICD-10 guide. The x-axis represents the predicted labels, while the y-axis represents the ground truth labels. The diagonal cells reflect the correct predictions. The intensity of the blue shading in each cell indicates the frequency of occurrences. Confusion matrix (b) compares the ICD-10 codes generated by the vanilla LLaMA2 with the ICD-10 codes from ground truth. Comparison of confusion matrix (a) and (b) shows that fine-tuning significantly enhanced LLaMA2's performance in ICD-10 code prediction. Similar results for the Mistral model are presented in Supplementary Figure 2.

# Discussion

Although our work was initiated with fine-tuning small-sized 7B LLMs for the feasibility study on radiation oncology tasks, it had attracted a lot of attention and interest from both the radiation oncology community and beyond. We appreciate the valuable support and insightful discussions received throughout our work. In this article, we have selected a few key topics for discussion due to the word limit. As the exploration of LLMs in radiation oncology continues growing, we will include more discussions in our future work.

## Importance of LLMs utilization in radiation oncology tasks

Radiation oncology integrates multiple disciplines like physics, biology, and medicine to develop personalized cancer treatment plans. It can significantly reduce the time consumed by radiation oncologists and improve work efficiency by automating the process radiation oncologists need to access the patient clinical notes and medical references to provide tailored treatment recommendations to patients. By fine-tuning LLMs to assist the patient treatment plan-making and predicting treatment modalities, radiation oncologists may be able to expedite the planning process, allowing for more timely initiation of treatment and improved patient outcomes. Accurate ICD code prediction can facilitate efficient billing and reimbursement processes, reducing administrative burdens on healthcare providers. Utilizing fine-tuned LLMs to help streamline the assignment of ICD codes will greatly improve the efficiency of record-keeping and administrative tasks. Our findings suggest that fine-tuned LLMs have great potential to make meaningful contributions to these tasks and support radiation oncologists in decision making.

## Data quality for fine-tuning

High-quality data is crucial for successful supervised fine-tuning. As evidenced in previous work, supervised fine-tuning annotations in the order of tens of thousands is enough to achieve high-quality results[14]. In light of this, we were motivated to collect and annotate 7,903 high-quality patient cases from our radiation oncology database. The emphasis on "high quality" means that each case was selected based on stringent criteria of only one corresponding primary treatment plan per patient diagnosis, and the annotation was performed with a high degree of accuracy and attention to detail. Validation was conducted on every patient to ensure completeness, accuracy, and consistency within the annotated data. Although we spent a lot of time collecting and curating the data, the benefit of using a large amount of high-quality data is significant. We have observed a great reduction in hallucination and mode failure during the fine-tuning process with our curated dataset, and a significant improvement in the performance of LLMs on the three radiation oncology tasks.

## Clinical evaluation and institutional practice

Despite well-defined grading definitions for evaluating generated treatment regimens, there are still a few evaluation cases that are difficult to grade, i.e. the model generating correct prescriptions but to a removed disease site. This kind of hallucination commonly seen in all LLMs although reduced during fine-tuning, still exists. An example case and more discussions are included in Supplementary 1.4. As it is extremely difficult to aggregate data at the scale needed from multiple institutions, our models were fine-tuned on a single-institutional dataset. But it is worth noting that our institutional practice closely adheres to the National Comprehensive Cancer Network (NCCN) guidelines, which are standardized frameworks for cancer treatment clinical decision making across institutes. Treatment choices can depend on patient-specific conditions, doctor-specific prescription habits, hospital resources, etc. Different treatment options might be clinically acceptable for the same patient, which poses extra challenges for fine-tuning LLMs. This study is not to provide universal solutions applicable to all institutions, but to demonstrate that fine-tuning LLMs with domain knowledge can significantly improve their performance in healthcare tasks. The generated answers still need to be interpreted by human experts with caution.

### Clinical impact of fine-tuned LLMs

In this study, we fine-tuned open-source LLMs for radiation oncology using the short diagnosis as input and observed a significant improvement in the performance of their output on the three clinical tasks. Our clinical evaluation revealed that over 60% of the treatment regimen generated by the fine-tuned LLMs is clinically acceptable. While these results are very promising, we are also aware that there may still exist a distance from applying the fine-tuned LLMs in general to real clinical scenarios, especially since we are using the small-sized 7B LLMs for our fine-tuning and the short diagnosis text as input. However, as a demonstration study, our results have inspired future work, which is to use the much more detailed and lengthy notes as well as the more advanced larger-sized LLMs for further fine-tuning to generate more clinically relevant outputs, which will assist radiation oncology professionals to improve the efficiency and reduce the workload in real-world scenarios.

### Fine-tuned small-sized LLMs vs vanilla large-sized LLMs

With the growth of AI, the performance of the vanilla LLMs on the market has been constantly improved, and the size of the LLMs is growing. For some easy text summarizing and extracting tasks, we have seen an acceptable performance by the vanilla large-sized LLMs. However, in the scope of this study focusing on the three radiation oncology clinical tasks, we still see the underperformance of the large-sized LLMs. For example, we have observed that the fine-tuned 7B LLaMA-2 model can still outperform the vanilla 70B LLaMA-3.1 model. This indicates that the domain-specific data, especially the curated high-quality data, used to fine-tune the small-sized LLM, may have an ineligible impact on the improvement of performance of the LLMs on the specific tasks. However, due to the scope of this study, more results including comparing fine-tuned small-sized LLMs with original large-sized LLMs on the radiation oncology tasks will be reported in our future work.

Limitations

Although our fine-tuned models demonstrated improved performance over the original models, we also noticed limitations in our study.

First, we still observed some mistakes made by the fine-tuned LLMs due to hallucinations, although we have observed a reduction in hallucinations while fine-tuning with our curated data. This is because LLMs predict the next tokens based on probabilities learned during training, and hallucinations are the inherent probabilistic nature of LLMs.

Second, the fine-tuned models only demonstrated improved performance on the specific tasks that the models were fine-tuned for. They did not exhibit capabilities beyond these tasks, as they were neither pre-trained for these tasks beyond. Thus, it may depend on different clinical tasks for the medical professionals to decide whether to go through the fine-tuning approach or not. Some LLMs may have performed very well on easy tasks like summarization, keyword extraction, etc. without fine-tuning. But for the very specific tasks, like the ones presented in this study. Fine-tuning was demonstrated to a way to improve the performance of LLMs on those tasks.

Thirdly, although our fine-tuned models have improved performance on clinical tasks in radiation oncology, we still noticed a distance when applying our fine-tuned LLMs to the clinic. For example, we used short diagnostic notes as input to fine-tune LLMs, while in the real-world scenario, patients often have much lengthy and detailed clinical notes. We also excluded patients with multiple diagnostic records or alternative primary treatment during the data curation, while these records and plans do exist in the clinic. As a demonstration study, we utilized 7B small LLMs for the fine-tuning, while larger size LLMs when fine-tuned often illustrate better performance. But, as one of the first steps of evaluating and trying to apply fine-tuned LLMs to radiation oncology clinical tasks, we understand these limitations and our findings demonstrated in this study have inspired us to advance further in adapting LLMs to assist radiation oncology tasks.

## Conclusion

We fine-tuned LLaMA-2 7B and Mistral 7B models with radiation oncology domain data and achieved statistically significant improvements in three clinical tasks: treatment regimen generation, treatment modality selection, and ICD-10 code prediction. The clinical evaluation on the treatment regimens generated by our fine-tuned models revealed that over 60% are clinically acceptable. As a feasibility study, although there is still a distance of applying our fine-tuned models directly to the clinic, it has demonstrated the great potential of LLMs in radiation oncology and inspired further development of LLMs for radiation oncology tasks.

## References

1.	Dawson LA, Jaffray DA. Advances in Image-Guided Radiation Therapy. Journal of Clinical Oncology. 2007;25(8):938-46. doi: 10.1200/jco.2006.09.9515. PubMed PMID: 17350942.
2.	Xing L, Thorndyke B, Schreibmann E, Yang Y, Li T-F, Kim G-Y, Luxton G, Koong A. Overview of image-guided radiation therapy. Medical Dosimetry. 2006;31(2):91-112. doi: https://doi.org/10.1016/j.meddos.2005.12.004.

3.	Liu C, Liu Z, Holmes J, Zhang L, Zhang L, Ding Y, Shu P, Wu Z, Dai H, Li Y, Shen D, Liu N, Li Q, Li X, Zhu D, Liu T, Liu W. Artificial general intelligence for radiation oncology. Meta-Radiology. 2023;1(3):100045. doi: https://doi.org/10.1016/j.metrad.2023.100045.
4.	Huynh E, Hosny A, Guthier C, Bitterman DS, Petit SF, Haas-Kogan DA, Kann B, Aerts HJWL, Mak RH. Artificial intelligence in radiation oncology. Nature Reviews Clinical Oncology. 2020;17(12):771-81. doi: 10.1038/s41571-020-0417-8.
5.	Achiam J, Adler S, Agarwal S, Ahmad L, Akkaya I, Aleman FL, Almeida D, Altenschmidt J, Altman S, Anadkat S. Gpt-4 technical report. arXiv preprint arXiv:230308774. 2023.
6.	Zhao L, Zhang L, Wu Z, Chen Y, Dai H, Yu X, Liu Z, Zhang T, Hu X, Jiang X, Li X, Zhu D, Shen D, Liu T. When brain-inspired AI meets AGI. Meta-Radiology. 2023;1(1):100005. doi: https://doi.org/10.1016/j.metrad.2023.100005.
7.	Eggmann F, Weiger R, Zitzmann NU, Blatz MB. Implications of large language models such as ChatGPT for dental medicine. J Esthet Restor Dent. 2023;35(7):1098-102. Epub 20230405. doi: 10.1111/jerd.13046. PubMed PMID: 37017291.
8.	Holmes J, Liu Z, Zhang L, Ding Y, Sio TT, McGee LA, Ashman JB, Li X, Liu T, Shen J, Liu W. Evaluating large language models on a highly-specialized topic, radiation oncology physics. Front Oncol. 2023;13:1219326. Epub 20230717. doi: 10.3389/fonc.2023.1219326. PubMed PMID: 37529688; PMCID: PMC10388568.
9.	Deng W, Ding X, Younkin JE, Shen J, Bues M, Schild SE, Patel SH, Liu W. Hybrid 3D analytical linear energy transfer calculation algorithm based on precalculated data from Monte Carlo simulations. Med Phys. 2020;47(2):745-52. Epub 2019/11/24. doi: 10.1002/mp.13934. PubMed PMID: 31758864.
10.	Shan J, Yang Y, Schild SE, Daniels TB, Wong WW, Fatyga M, Bues M, Sio TT, Liu W. Intensity-modulated proton therapy (IMPT) interplay effect evaluation of asymmetric breathing with simultaneous uncertainty considerations in patients with non-small cell lung cancer. Med Phys. 2020;47(11):5428-40. Epub 2020/09/24. doi: 10.1002/mp.14491. PubMed PMID: 32964474; PMCID: PMC7722083.
11.	Zaghian M, Cao W, Liu W, Kardar L, Randeniya S, Mohan R, Lim G. Comparison of linear and nonlinear programming approaches for "worst case dose" and "minmax" robust optimization of intensity-modulated proton therapy dose distributions. J Appl Clin Med Phys. 2017;18(2):15-25. Epub 2017/03/17. doi: 10.1002/acm2.12033. PubMed PMID: 28300378; PMCID: PMC5444303.
12.	Deng W, Yang Y, Liu C, Bues M, Mohan R, Wong WW, Foote RH, Patel SH, Liu W. A Critical Review of LET-Based Intensity-Modulated Proton Therapy Plan Evaluation and Optimization for Head and Neck Cancer Management. Int J Part Ther. 2021;8(1):36-49. Epub 2021/07/22. doi: 10.14338/IJPT-20-00049.1. PubMed PMID: 34285934; PMCID: PMC8270082.
13.	Schild SE, Rule WG, Ashman JB, Vora SA, Keole S, Anand A, Liu W, Bues M. Proton beam therapy for locally advanced lung cancer: A review. World J Clin Oncol. 2014;5(4):568-75. Epub 2014/10/11. doi: 10.5306/wjco.v5.i4.568. PubMed PMID: 25302161; PMCID: PMC4129522.
14.	Touvron H, Martin L, Stone K, Albert P, Almahairi A, Babaei Y, Bashlykov N, Batra S, Bhargava P, Bhosale S, Bikel D, Blecher L, Canton Ferrer C, Chen M, Cucurull G, Esiobu D, Fernandes J, Fu J, Fu W, Fuller B, Gao C, Goswami V, Goyal N, Hartshorn A, Hosseini S, Hou R, Inan H, Kardas M, Kerkez V, Khabsa M, Kloumann I, Korenev A, Singh Koura P, Lachaux M-A, Lavril T, Lee J, Liskovich D, Lu Y, Mao Y, Martinet X, Mihaylov T, Mishra P, Molybog I, Nie Y, Poulton A, Reizenstein J, Rungta R, Saladi K, Schelten A, Silva R, Smith EM, Subramanian R, Tan XE, Tang B, Taylor R, Williams A, Kuan JX, Xu P, Yan Z, Zarov I, Zhang Y, Fan A, Kambadur M, Narang S, Rodriguez A, Stojnic R, Edunov S, Scialom T. Llama 2: Open Foundation and Fine-Tuned Chat Models2023 July 01, 2023:[arXiv:2307.09288 p.].
15.	Jiang AQ, Sablayrolles A, Mensch A, Bamford C, Singh Chaplot D, de las Casas D, Bressand F, Lengyel G, Lample G, Saulnier L, Renard Lavaud L, Lachaux M-A, Stock P, Le Scao T, Lavril T, Wang T, Lacroix T, El Sayed W. Mistral 7B2023 October 01, 2023:[arXiv:2310.06825 p.].

16.	Hu EJ, Shen Y, Wallis P, Allen-Zhu Z, Li Y, Wang S, Wang L, Chen W. LoRA: Low-Rank Adaptation of Large Language Models2021 June 01, 2021:[arXiv:2106.09685 p.].